\documentclass[12pt]{iopart}
\usepackage{psfrag,epsfig}
\usepackage{graphicx}
\newcommand{\vp}{\varphi}

\newcommand{\ve}{\varepsilon}

\newcommand{\kB}{k_{\mathrm{B}}}
\newcommand{\hot}{{\mathrm{h}}}
\newcommand{\cold}{{\mathrm{c}}}
\newcommand{\gcc}{\langle g_\cold^2 \rangle}
\newcommand{\ghh}{\langle g_\hot^2 \rangle}
\newcommand{\gccc}{\langle g_\cold^3 \rangle}
\newcommand{\ghhh}{\langle g_\hot^3 \rangle}

\begin{document}
\title{Energetics of a Microscopic Feynman Ratchet}
\author{Bart Cleuren$^1$ and Ralf Eichhorn$^2$}
\address{$^1$UHasselt, Faculty of Sciences, Theory Lab, Agoralaan, 3590 Diepenbeek, Belgium}
\address{$^2$Nordita, Royal Institute Technology and Stockholm University,\\Hannes Alfv\'ens v\"ag 12 , SE-106 91 Stockholm, Sweden}
\eads{bart.cleuren@uhasselt.be, eichhorn@nordita.org}

\begin{abstract} 
A general formalism is derived describing both dynamical and energetic properties of a microscopic Feynman ratchet. Work and heat flows are given as a series expansion in the thermodynamic forces, obtaining analytical expressions for the (non)linear response coefficients. Our results extend previously obtained expressions in the context of a chiral heat pump.
\end{abstract}

\submitto{\JSTAT}
\date{\today}
\tableofcontents
\maketitle
\section{Introduction}
The Feynman ratchet is a celebrated example of a thermal engine \cite{::Feynman}. Designed to illustrate the second law of thermodynamics, it is capable to extract useful work from a heat flow between two thermal reservoirs at different temperatures. The engine consists of two rigidly connected parts, a ratchet and pawl at one end of a rotational axle, and vanes at the other end. Each part is immersed in a surrounding gas, and collisions between the gas particles and the device cause a random rotational motion. The shape of the ratchet introduces a rotational asymmetry which leads to a systematic rotation in a certain direction \cite{Reimann:PR2002:Brownian}, and is driven by the exchange of heat from the hot to the cold reservoir. Using this rotation for example to lift a weight against the force of gravity, the whole setup can deliver useful work. On the basis of hand-waving arguments, Feynman claims in his \emph{Lectures on Physics} that the engine is capable of operating at Carnot efficiency. It was realised later \cite{Parrondo:AJoP1996:Criticism} that the efficiency inevitably has to be lower, as the engine is at all times in simultaneous contact with the two heat reservoirs, resulting in a heat leakage.

A microscopic analysis of the dynamics and energetics of the original engine has proven to be notoriously difficult.  In part this is caused by the unavoidable recollisions of the gas particles with the engine giving rise to correlations. Various approaches have been considered to circumvent these difficulties, for example by modelling the collisions by Langevin noise (eg. \cite{Parrondo:AJoP1996:Criticism,Magnasco1998}) or by considering a discrete setup, eg. \cite{PhysRevE.59.6448,PhysRevE.106.014154}. An alternative approach was developed by Van den Broeck and co-workers in a series of papers in which they meticulously stripped down the engine to its basic constitutes \cite{Broeck:NCSSD2003:Microscopic,VandenBroeck:PRL2004:Microscopic,Meurs:PRE2004:Rectification,Broeck:NJP2005:Maxwell,Meurs:JPCM2005:Thermal}, see also \cite{feigel:PRE2017}. Considering ideal gases and convex engine parts (which replace the vanes, ratchet and pawl) an exact microscopic description is derived which is centred around the stochastic time evolution of the (angular) velocity of the device. Analytical expressions for the average velocity are obtained in the form of a series expansion in $m/M$, the mass ratio between gas particles and device. Even in such an ideal situation, the interplay between the geometry and the collisions is intricate and it is not at all obvious in which direction the engine will turn. A description of the energetics was obtained by adding a torque \cite{VandenBroeck:PRL2006:Brownian,chiralFridge}. In those papers the work and heat flows were derived from expressions of the average velocity in the linear regime and making use of the Onsager symmetry. The purpose of the present work is to go beyond the linear regime by setting the heat and work variables on the same footing as the angular velocity. Such an extended framework allows to calculate the average (and higher moments of) work and heat up to any order of the thermodynamic forces.

The paper is organized as follows. Section \ref{sec:MFR} introduces the microscopic Feynman ratchet and describes the dynamics of the engine as the result of the external torque and collisions with the gas particles. Section \ref{sec:EAT} extends this framework by incorporating both work and heat variables. The main results are presented in section \ref{sec:R}, clarifying the influence of the geometry on the motion of the rotor. Finally, we briefly conclude in section~\ref{sectConc}. 
\begin{figure}[t]
\begin{center}
\includegraphics[width=0.8\columnwidth]{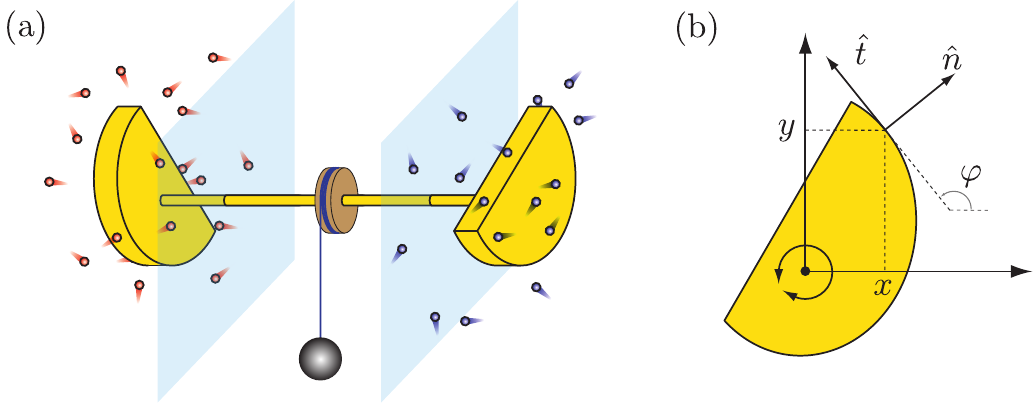}
\end{center}
\caption{(a) Sketch of the system under consideration: the engine consists of two convex objects rigidly connected by an axle, which serves as the fixed rotational axis. Mounted along the axle is a weight, which is lifted as a result of the collisions of the gas particles with the convex objects. (b) Each convex object is homogeneous along the rotational axis, which allows for a two dimensional analysis. Shown is a top view of the object, indicating the geometrical parameters necessary to describe the collisions with the gas particles.}
\label{fig1}
\end{figure}

\section{Microscopic Feynman Ratchet}
\label{sec:MFR}
The original Feynman ratchet was constructed by using a ratchet and pawl on one end of a rigid rotational axle and a set of vanes on the other end. Because of the spatial complexity of such a construction, colliding gas particles are very likely to collide with the various engine parts more than once. This introduces correlations, and hence a memory, making analytical calculations overwhelmingly complicated. In order to eliminate such recollisions, and in effect make individual collisions to occur independently, we reduce the setup to its essence \cite{Broeck:NCSSD2003:Microscopic}. The purpose of the vanes is to allow transfer of energy from the gas particles to the engine and vise versa. The ratchet and pawl set-up introduces a spatial asymmetry which ensures a difference between clockwise/anti-clockwise rotation of the engine. A conceptually simple construction with these properties is shown in figure \ref{fig1}(a). It consists of two objects rigidly connected by the rotational axis. For such a construction recollisions can be reduced strongly (so that they can be neglected in a subsequent analysis) by requiring (i) the shape of each object to be convex and (ii) the total mass $M$ of the engine to be much larger than the mass $m$ of the gas particles. The first requirement is purely geometrical and ensures that a gas particle is directed away from the object after colliding. The second requirement ensures that a particle, again after collision, is not caught up by the engine. Each object is surrounded by an ideal gas at a certain temperature and density. In the context of a heat engine one of the surrounding ideal gases is labeled as the cold reservoir (sub- or superscript $\cold$ in the calculations below) and the other one is then the hot reservoir (sub- or superscript $\hot$). A rotational asymmetry is implemented by shaping the objects asymmetrically with respect to their points of rotation (see Fig.\ref{fig1}(b)), and by arranging them in an asymmetric fashion with respect to each other. Finally the last ingredient is a weight mounted on the rotational axle which exerts a constant torque $\tau$.

By construction, the whole engine has only one degree of freedom, namely the rotation around the connecting axle. Its dynamical state is thus fully described by the instantaneous angular velocity $\omega$. In time this velocity changes either due to random collisions of gas particles with any of the two convex objects, or due to the action of the external torque. The effect of the latter is deterministic and described by
\begin{equation}\label{eq:domegadt}
    I\frac{d \omega}{dt}=\tau
\end{equation}
where $I$ is the total moment of inertia around the rotational axis of the whole engine.

Quantifying the effect of collisions is more elaborate. Before we start let us set the scene as follows. The surfaces of the convex objects and the gas particles are taken to be smooth, so no tangential forces occur during collisions. As a consequence, collisions of gas particles with the objects' sides which are oriented perpendicular to the connecting axle do not affect $\omega$. Additionally we take the convex rotors to not vary in shape along the direction of the rotational axis, as seen in Fig.~\ref{fig1}(a). Hence collisions with the engine can be analysed in an effective two-dimensional setup. In order to calculate the effect of a collision, we only need to consider a two-dimensional convex object as sketched in Fig.~\ref{fig1}(b). It is free to rotate around a fixed axis, which is orthogonal to the plane of the object. In that plane we choose a fixed coordinate system with origin located a the center of rotation such that the $z$-axis coincides with the rotational axis. The orientation of the $x$ and $y$ axis is arbitrary. The instantaneous angular velocity is therefore $\vec{\omega}=\omega \vec{e}_{z}$. The velocity $\vec{V}$ of an arbitrary point $\vec{r}=(x,y)$ on the surface
of the object is $\vec{V}=\vec{\omega} \times \vec{r}=\omega (-y,x)$. The orientation of the surface at a point $\vec{r}$ is specified by the angle $\vp$ between the $x$-axis and the tangential unit vector
$\hat{t}=(\cos \vp, \sin \vp)$. The normal to the surface
pointing outward is $\hat{n}=(\sin \vp, -\cos \vp)$.

We now consider a gas particle with mass $m$ and
(pre-collisional) speed $\vec{v}=(v_{x},v_{y})$ as it
hits the surface of one of the objects $\alpha\in\{\cold,\hot\}$
at position $\vec{r}=(x,y)$ which moves with
(pre-collisional) velocity $\omega (-y,x)$. For analyzing this collision we need to consider the \emph{total} mass $M$ and \emph{total} moment of inertia $I$ around the rotational axis of the engine as a whole, i.e.\ of \emph{both} convex rotors taken together, because they are rigidly connected. The post-collisional velocities (labeled with a prime $'$ within this Section) can be calculated
on the basis of the following conservation laws \cite{chiralFridge,Cleuren:JSM2008:Dynamical}. First, taking the collisions to occur instantaneously, so that the external torque does not affect $\omega$ during the collision, the angular momentum along the $z$-axis is conserved,
$m(x v_y'-y v_x')+I\omega'=m(x v_y-y v_x)+I\omega$.
Second, as there are no tangential forces the tangential velocity of the gas particle is conserved, $\vec{v}'\cdot\hat{t}=\vec{v}\cdot\hat{t}$,
while the normal component of the relative velocity changes sign,
$(\vec{v}'-\vec{V}')\cdot\hat{n}=-(\vec{v}-\vec{V})\cdot\hat{n}$.
These three conditions uniquely determine the post-collisional velocities,
\numparts
\begin{eqnarray}
\omega' = \omega +\frac{2\ve^2g_\alpha(\vec{V}-\vec{v})\cdot \hat{n}}{R_I(1+\ve^2 g_\alpha^{2})}
\label{eq:CRomega}
\, , \\
v_{x}' = v_{x}+\frac{2\sin\vp\,(\vec{V}-\vec{v})\cdot \hat{n}}{1+\ve^2 g_\alpha^{2}}
\, , \quad
v_{y}' = v_{y}-\frac{2\cos\vp\,(\vec{V}-\vec{v})\cdot \hat{n}}{1+\ve^2 g_\alpha^{2}}
\label{eq:CRvxvy}
\end{eqnarray}
\endnumparts
for which the following quantities are introduced:
\numparts
\begin{eqnarray}
\ve = \sqrt{m/M}
\label{eq:eps}
\, , \\
R_I = \sqrt{I/M}
\label{eq:RI}
\, , \\
g_\alpha \equiv g_\alpha(x,y,\vp)
= \frac{x}{R_I}\cos \vp+\frac{y}{R_I}\sin \vp
\label{eq:galpha}
\, .
\end{eqnarray}
\endnumparts
Note that the first two parameters, the mass ratio $\ve$ and the radius of inertia $R_I$, are global properties of the engine, whereas the third quantity $g_\alpha$ contains information on the geometry (see Fig.~\ref{fig1}(b)) of the specific part of the engine which is hit by the gas particle in one of the reservoirs and therefore carries the subscript $\alpha\in\{\cold,\hot\}$.

Knowledge of the post-collisional velocities in function of the pre-collisional ones allows to derive the probability per unit time $W_\alpha(\omega'\vert\omega)$ for the angular velocity to change from $\omega$ to $\omega'$ due to collisions of gas particles with the convex object in reservoir $\alpha\in\{\cold,\hot\}$:
\begin{eqnarray}
W_\alpha(\omega'\vert\omega) =
\int \rmd S_\alpha \int \rmd v_x \int \rmd v_y \, \rho_{\alpha}\phi_\alpha(v_x,v_y) 
\, \Theta\left[(\vec{V}-\vec{v})\cdot\hat{n}\right]
\nonumber \\
\qquad\qquad\qquad\qquad
\times \left|(\vec{V}-\vec{v})\cdot\hat{n}\right| \,
\delta\left[ \omega'-\omega - \frac{2\ve^2g_\alpha(\vec{V}-\vec{v})\cdot \hat{n}}{R_I(1+\ve^2 g_\alpha^{2})} \right]
\, .
\end{eqnarray}
This expression is explained as follows. First there is the integral over the object's one-dimensional surface $S_\alpha$ which adds up the contributions
from all possible points of impact. For each point of impact, the Dirac-$\delta$ function picks out those velocities $\vec{v}$ of the incoming gas particles which lead to the required change $\omega'-\omega$ in angular velocity. The Heaviside-function $\Theta$ ensures that the gas particle is moving towards the rotating
object. The likelihood for a collision with that specific velocity depends on the properties of the ideal gas in the reservoir $\alpha\in\{\cold,\hot\}$ which enter the equation via the particle density $\rho_\alpha$ and the Maxwell-Boltzmann distribution $\phi_\alpha(v_x,v_y)$
at temperature $T_\alpha$. As this distribution is Gaussian the velocity integrals can be done explicitly and one obtains the expression
\begin{eqnarray}\label{tr}
W_\alpha(\omega'\vert\omega) = 
\frac{\rho_\alpha}{4}\sqrt{\frac{I^2}{2\pi m \kB T_\alpha}}
\int \rmd S_\alpha \, \Theta\left[\frac{\omega'-\omega}{g_\alpha}\right] \left\vert \omega'-\omega \right\vert
\frac{(1+\ve^2 g_\alpha^2)^2}{\ve^2 g_\alpha^2}
\nonumber \\
\qquad\qquad\qquad\qquad
\times \exp\left[
	-\frac{I}{2\kB T_\alpha} \left(\omega \ve g_\alpha + \frac{(\omega'-\omega)(1+\ve^2 g_\alpha^2)}{2\ve g_\alpha}\right)^2
\right]\, .
\end{eqnarray}
Finally, since the collisions in the hot and cold gas reservoir occur independently of each other, the contributions from the two convex objects simply add up, such that the total rate for transitions of the angular velocity from $\omega$ to $\omega'$ is $W_\cold(\omega'\vert\omega) + W_\hot(\omega'\vert\omega)$.

Having established the transition rates, cf Eq.~\ref{tr}, together with the deterministic evolution due to the torque given by Eq.~\ref{eq:domegadt}, it is straightforward to write down the Master equation describing the time evolution of the probability density $P(t;\omega)$
\begin{eqnarray}\label{me_omega}
\frac{\partial P(t;\omega)}{\partial t} & = &
-\frac{\tau}{I}\frac{\partial P(t;\omega)}{\partial \omega}
+\int \rmd\omega' \left[ W_\cold(\omega\vert\omega') + W_\hot(\omega\vert\omega') \right] P(t;\omega') \nonumber \\
&& \qquad\qquad\qquad\mbox{} - \int \rmd\omega' \left[ W_\cold(\omega'\vert\omega) + W_\hot(\omega'\vert\omega) \right] P(t;\omega)\,.
\end{eqnarray}
Instead of solving this equation for $P(t;\omega)$ directly, we use a different approach by focusing on the moments of the distribution $P(t;\omega)$,
\begin{equation}
    \langle \omega^n \rangle \equiv \int \rmd\omega \;\omega^n P(t;\omega)
\, .
\end{equation}
Using these moments, the master equation can be transformed to the equivalent (infinite) set of evolution equations \cite{feigel:PRE2017,Cleuren:JSM2008:Dynamical}:
\begin{equation}\label{eq_moments}
\frac{\rmd}{\rmd t}\langle \omega^n \rangle = \frac{n \tau}{I}\langle \omega^{n-1}\rangle + \sum_{k=1}^{n} {n\choose k} \left\langle \omega^{n-k} \left[ a_k^{(\cold)}(\omega)+a_k^{(\hot)}(\omega) \right] \right\rangle
\, .
\end{equation}
Here we introduced the so-called jump moments \begin{equation}
\label{eq:an}
a_n^{(\alpha)}(\omega) \equiv \int \rmd\omega' \, (\omega'-\omega)^n \, W_\alpha(\omega'\vert \omega)
\end{equation}
for the each of the reservoirs $\alpha \in \{\cold,\hot\}$.
The equations for the first two moments are
\numparts
\begin{eqnarray}
\frac{\rmd}{\rmd t}\langle \omega \rangle
	= \frac{\tau}{I} + \langle a_1^{(\cold)}(\omega) \rangle + \langle a_1^{(\hot)}(\omega) \rangle
\label{eq:d<omega>dt}
\, , \\[1ex]
\frac{\rmd}{\rmd t}\langle \omega^2 \rangle
	= \frac{2\tau}{I} \langle \omega \rangle
	+ 2 \left\langle \omega \left[ a_1^{(\cold)}(\omega) + a_1^{(\hot)}(\omega) \right]  \right\rangle
	+ \langle a_2^{(\cold)}(\omega) \rangle + \langle a_2^{(\hot)}(\omega) \rangle
\, .
\label{eq:d<omega2>dt}
\end{eqnarray}
\endnumparts
Since there is no approximation involved in deriving \eref{eq_moments}, these set of equations is fully equivalent to the master equation \eref{me_omega}, and is equally difficult to solve analytically. We thus solve \eref{eq:d<omega>dt}, \eref{eq:d<omega2>dt} only in the stationary limit, and resort to using an expansion in the small parameter $\varepsilon=\sqrt{\frac{m}{M}}$. When performing this expansion, we have to bear in mind that in the stationary state fluctuations of $\omega$ are of the order of $\sqrt{\kB T/I} \sim \sqrt{1/M}$ (with an average temperature $T$, see below), i.e.\ we have to perform the expansion for the rescaled quantity $\sqrt{M}\omega$ \cite{Cleuren:JSM2008:Dynamical}. Moreover, we have to rescale the torque $\tau$ by $M/m$ to find a stationary state with finite average rotational speed $\langle\omega\rangle$, resulting from a balance between external torque and frictional forces. Specifically, we find
\begin{eqnarray}
\langle \omega \rangle &=&
\frac{\tau}{\gamma_\cold+\gamma_\hot}
+ \frac{\tau^2}{(\gamma_\cold+\gamma_\hot)^2}
	\frac{m R_I^3
	 	  \left( \langle g_\cold^3 \rangle \rho_\cold + \langle g_\hot^3 \rangle \rho_\hot \right)} 
	 	 {\gamma_\cold+\gamma_\hot}
\nonumber\\&& \qquad \mbox{}
+ \frac{m}{M}\kB(T_\hot-T_\cold)
	\frac{R_I \left( \gamma_\hot \langle g_\cold^3 \rangle \rho_\cold
		- \gamma_\cold \langle g_\hot^3 \rangle \rho_\hot \right)}
		 {(\gamma_\cold+\gamma_\hot)^2}
+ \ldots
\end{eqnarray}
and
\begin{eqnarray}
    \langle \omega^2 \rangle &=& \frac{\tau^2}{\left(\gamma_c+\gamma_h\right)^2}+\frac{k}{I}\frac{\gamma_c T_\cold+\gamma_h T_\hot}{\gamma_\cold+\gamma_\hot}+\ldots
\end{eqnarray}
where we introduced the friction coefficients
\begin{equation}
    \gamma_{\alpha}= 2\rho_{\alpha}\sqrt{\frac{2m\kB T_{\alpha}}{\pi}}R_I^2 \langle g_{\alpha}^2\rangle
\, .
\end{equation}
The symbol $\langle g_\alpha^k \rangle$ (for $k=1,2,3,\ldots$) denotes a average of $g_\alpha^k$,
defined in Eq.\ \ref{eq:galpha},
over the surface of the object in reservoir $\alpha$%
\footnote{Note that this definition differs from the definition we used in \cite{Cleuren:JSM2008:Dynamical} by a factor of unit length representing the total ``surface'' of the object.},
\begin{equation}
   \langle g_\alpha^k \rangle
   = \int \rmd S_\alpha \, g_\alpha^k
   = \int \rmd S_\alpha \left( \frac{x}{R_I}\cos \vp+\frac{y}{R_I}\sin \vp \right)^{\!k}
\, .
\end{equation}
\section{Energetics and thermodynamics: general framework} \label{sec:EAT}
Having established the dynamics of the angular velocity $\omega$, via the master equation and corresponding transition rates, we now add the thermodynamic quantities into the framework. In the context of the ratchet system from Fig.~\ref{fig1} these quantities are heat $Q_\cold$ and $Q_\hot$ and work $W$. We define them as positive if delivered to the system. All energy exchanges are mediated by the rotational movement of the engine, and the conceptual simplicity makes a clean distinction between heat and work straightforward. At a momentary rotational speed $\omega$, the rate of work (power) delivered by the external torque is given by
\begin{equation}\label{eq:dWdt}
    \frac{\rmd W}{\rmd t}=\tau \omega\, .
\end{equation}
The exchange of heat with the reservoirs is due to the instantaneous collisions of the engine with the gas particle. As each collision is elastic, the kinetic energy change of the imparting gas particle is fully transferred to the rotational energy of the engine. Hence, upon a collision with a gas particle in reservoir $\alpha$ $\left(\in\{\cold,\hot\}\right)$  that changes the angular velocity from $\omega$ to $\omega'$, the amount of heat received from the reservoir is given by the corresponding change of kinetic energy
\begin{equation} \label{eq:DeltaQalpha}
    \Delta Q_\alpha = \frac{I}{2} \left[ (\omega')^{2}-\omega^{2} \right]\, .
\end{equation}
It is clear from these definitions that work and heat are fluctuating random variables, as they depend on $\omega$. Our main goal is to analyze the thermodynamic properties of the engine by calculating the lowest order moments $\langle Q_\cold \rangle$, $\langle Q_\hot \rangle$, and $\langle W \rangle$.

The first step is to incorporate the new quantities as extra variables into the Master equation. The full state is then given by the combined set of variables $\omega$, $W$ and $Q_\alpha$, where $\alpha\in\{\cold,\hot\}$ can be chosen to represent either of the reservoirs. The time evolution of the probability density $P(t;\omega,W,Q_\alpha)$ is
\begin{eqnarray}
\frac{\partial P(t;\omega,W,Q_\alpha)}{\partial t} & = &
-\frac{\tau}{I}\frac{\partial P(t;\omega,W,Q_\alpha)}{\partial \omega}-\tau \omega\frac{\partial P(t;\omega,W,Q_\alpha)}{\partial W}
\nonumber \\ && \quad\mbox{}
+\int\rmd\omega'\, W_\alpha(\omega\vert\omega')P(t;\omega',W,Q_\alpha-\Delta Q_\alpha)
\nonumber \\ && \quad\mbox{}
+\int\rmd\omega'\, W_{\bar\alpha}(\omega\vert\omega')P(t;\omega',W,Q_\alpha)
\nonumber \\ && \quad\mbox{}
-\int\rmd\omega' \left[ W_\cold(\omega'\vert\omega) + W_\hot(\omega'\vert\omega) \right] P(t;\omega,W,Q_\alpha)
\label{eq:ME}
\, .
\end{eqnarray}
Compared to Eq.~\ref{me_omega} there is an extra term related to the (deterministic) evolution of $W$. Moreover, upon a collision one has to keep track of the reservoir, i.e.\ a collision in reservoir $\alpha$ induces a corresponding change of $Q_{\alpha}$, while collisions in the other reservoir (denoted by $\bar\alpha$) only change $\omega$. Integrating out the $W$ and $Q_\alpha$ dependence leads 
to Eq.~\ref{me_omega}.

As before, instead of solving the master equation, we turn our attention directly to the moments. The time-evolution of any moment of the quantities of interest $\omega$, $W$ and $Q_\alpha$ can be calculated according to
\begin{equation}
\langle f(\omega,W,Q_\alpha) \rangle = \int \rmd\omega \rmd W \rmd Q_\alpha \, f(\omega,W,Q_\alpha) P(t;\omega,W,Q_\alpha)
\, ,
\end{equation}
where the function $f$ denotes an arbitrary combination of its arguments. The evolution equations for the lowest order moments of work and heat are then
\begin{equation}
\frac{\rmd}{\rmd t}\langle W \rangle
	= \tau \langle \omega \rangle
\label{eq:d<W>dt}
\end{equation}
and
\begin{equation}
\frac{\rmd}{\rmd t}\langle Q_\alpha \rangle
	= I \langle \omega a_1^{(\alpha)}(\omega) \rangle + \frac{I}{2} \langle a_2^{(\alpha)}(\omega) \rangle
\label{eq:d<Qalpha>dt}
\, .
\end{equation}
A straightforward calculation yields
\begin{equation}
\frac{\rmd}{\rmd t}\langle W+Q_\hot+Q_\cold \rangle = \frac{\rmd}{\rmd t}\frac{I\langle \omega^2 \rangle}{2}
\, ,
\end{equation}
which simply expresses conservation of energy. One notices that the equation for the heat depends only on the jump moments of $\omega$. This is a direct consequence of the fact that the stochasticity of the heat $Q_\alpha$ is determined solely by the stochasticity of $\omega$.
The expression in Eq.~\ref{eq:d<Qalpha>dt} is obtained from the master equation, Eq.~\ref{eq:ME}, by multiplying it with $Q_\alpha$ and integrating over all variables $\omega$, $Q_\alpha$, $W$. This reduces the equation to $\rmd \langle Q_\alpha \rangle/\rmd t = \int\rmd\omega\rmd\omega'\, \Delta Q_\alpha W_\alpha(\omega'|\omega)P(t;\omega)$, where $\Delta Q_\alpha$ depends on both $\omega$ and  $\omega'$ according to Eq.~\ref{eq:DeltaQalpha}. In order to express this integral in terms of the jump moments, Eq.~\ref{eq:an}, we rewrite $\Delta Q_\alpha$ in powers of $(\omega'-\omega)$, i.e.~$(\omega')^2-\omega^2 = (\omega'-\omega)^2 + 2\omega(\omega'-\omega)$; from this, Eq.~\ref{eq:d<Qalpha>dt} directly follows.

\section{Energetics and thermodynamics: results and discussion} \label{sec:R}
The explicit expressions for work and heat quickly become unwieldy and, apart from the already introduced $\varepsilon$ expansion, we consider an additional expansion in terms of the so-called thermodynamics forces. These forces appear naturally when looking at the entropy production. From now on we focus on the stationary operation of the machine for which $\langle \dot{W} + \dot{Q}_\hot  +  \dot{Q}_\cold \rangle=0$, where the dot denotes the time-derivative.\newline
The entropy production in the reservoirs is given by the combined heat dissipation
\begin{equation}
\langle \Delta \dot S \rangle
	= -\frac{\langle \dot{Q}_\hot \rangle}{T_\hot} - \frac{\langle \dot{Q}_\cold \rangle}{T_\cold}\, .
\end{equation}
Eliminating one of the heat flows in favor of the work (cf conservation of energy) and noting that $\langle \dot{W} \rangle = \tau \langle \omega \rangle $ allows to reformulate this entropy production into the following two alternatives
\begin{equation}
\langle \Delta \dot S \rangle
	= \langle \omega \rangle \frac{\tau}{T_\hot}
		- \langle \dot{Q}_\cold \rangle \left( \frac{1}{T_\cold} - \frac{1}{T_\hot} \right)=\langle \omega \rangle \frac{\tau}{T_\cold}
		+ \langle \dot{Q}_\hot \rangle \left( \frac{1}{T_\cold} - \frac{1}{T_\hot} \right),
\end{equation}
There is \emph{a priori} no preference to choose one over the other, and so we choose to symmetrize by taking the arithmetic mean 
\begin{equation}
\langle \Delta \dot S \rangle
	= \langle \omega \rangle \frac{\tau}{2}\left(\frac{1}{T_\hot}+\frac{1}{T_\cold}\right)+\frac{\langle \dot{Q}_\hot \rangle- \langle \dot{Q}_\cold \rangle}{2} \left( \frac{1}{T_\cold} - \frac{1}{T_\hot} \right).
\end{equation}
This expression is easily cast into the familiar bilinear form
\begin{equation}
    \langle \Delta \dot S \rangle=J_1 X_1 + J_2 X_2
\end{equation}
after identifying the two thermodynamic forces
\begin{equation}
X_1 = \frac{\tau}{2}\left(\frac{1}{T_\hot}+\frac{1}{T_\cold}\right)
\, , \qquad
X_2 = \frac{1}{T_\cold} - \frac{1}{T_\hot}
\, ,
\end{equation}
and fluxes
\begin{equation}
J_1 =\langle \omega \rangle
\, , \qquad
J_2 = \frac{\langle \dot{Q}_\hot \rangle-\langle \dot{Q}_\cold \rangle}{2}
\, .
\end{equation}
As both fluxes depend on the thermodynamic forces, a series expansion leads in lowest order to the well-known Onsager coefficients
\begin{eqnarray}
J_1 = L_{11}X_1+L_{12}X_2
\, , \\
J_2 = L_{21}X_1+L_{22}X_2
\, .
\end{eqnarray}
The off-diagonal elements fulfil the Onsager symmetry $L_{12}=L_{21}$.\newline
Introducing the reference temperature
\begin{equation}
    \frac{1}{T}=\frac{1}{2}\left(\frac{1}{T_\hot}+\frac{1}{T_\cold}\right)
\end{equation}
allows to substitute $\tau$, $T_\hot$ and $T_\cold$ as
\begin{equation}
    \tau = X_1 T
    \, , \qquad
    \frac{1}{T_\hot}=\frac{1}{T}-\frac{X_2}{2}
    \, , \qquad
    \frac{1}{T_\cold}=\frac{1}{T}+\frac{X_2}{2}
    .
\end{equation}
Finally we are in a position to give the expressions for work and heat in terms of the relevant thermodynamic forces. Bear in mind that these are obtained by a series expansion in \emph{both} $\varepsilon$ and the thermodynamic forces $X_1$ and $X_2$. For the work we find
\numparts
\begin{eqnarray}
    \langle \dot{W} \rangle &=&
    \frac{1}{v m R_I^2}
\frac{T^2 X_1^2}{\langle g_\cold^2 \rangle \rho_\cold + \langle g_\hot^2 \rangle \rho_\hot}
\nonumber \\&& \mbox{}
+  \frac{\kB T}{v I} \rho_\cold \rho_\hot
	\frac{R_I\left(\langle g_\cold^3 \rangle \langle g_\hot^2 \rangle - \langle g_\cold^2 \rangle \langle g_\hot^3 \rangle \right)}
	     {\left( \langle g_\cold^2 \rangle \rho_\cold + \langle g_\hot^2 \rangle \rho_\hot \right)^2}
T^2 X_1 X_2
+ \ldots
\end{eqnarray}
The expressions for the heat are
\begin{eqnarray}
    \langle \dot{Q_\cold} \rangle &=&
- \frac{\kB T}{v I}
\rho_\cold \rho_\hot
	\frac{\langle g_\cold^3 \rangle \langle g_\hot^2 \rangle - \langle g_\cold^2 \rangle \langle g_\hot^3 \rangle}
	     {\left( \langle g_\cold^2 \rangle \rho_\cold + \langle g_\hot^2 \rangle \rho_\hot \right)^2}
T X_1
\nonumber \\&& \mbox{}
- \frac{m}{M} v \kB T
\frac{\langle g_\cold^2 \rangle \langle g_\hot^2 \rangle \rho_\cold \rho_\hot}
	 {\langle g_\cold^2 \rangle \rho_\cold + \langle g_\hot^2 \rangle \rho_\hot}
T X_2
\nonumber \\&& \mbox{}
+ \frac{\kB T}{4vI}
\rho_\cold \rho_\hot  T^2 X_1 X_2
\nonumber \\&& \mbox{} \times
\frac{R_I \left[\langle g_\cold^2 \rangle \rho_\cold \left( 19 \langle g_\cold^2 \rangle \langle g_\hot^3 \rangle
													- 5 \langle g_\hot^2 \rangle \langle g_\cold^3 \rangle \right)
	- \langle g_\hot^2 \rangle \rho_\hot \left( \langle g_\cold^2 \rangle \langle g_\hot^3 \rangle
													- 15 \langle g_\hot^2 \rangle \langle g_\cold^3 \rangle \right) \right]}
	 {\left( \langle g_\cold^2 \rangle \rho_\cold + \langle g_\hot^2 \rangle \rho_\hot \right)^3}
\nonumber \\&& \mbox{}
- \frac{1}{v m R_I^2}
	\frac{\langle g_\cold^2 \rangle \rho_\cold}
		 {\left( \langle g_\cold^2 \rangle \rho_\cold + \langle g_\hot^2 \rangle \rho_\hot \right)^2}
T^2 X_1^2
\nonumber \\&& \mbox{}
- \frac{v\kB T}{4}\frac{m}{M}
\frac{\langle g_\cold^2 \rangle \langle g_\hot^2 \rangle \rho_\cold \rho_\hot
		\left( \langle g_\cold^2 \rangle \rho_\cold - \langle g_\hot^2 \rangle \rho_\hot \right)}
	 {\left( \langle g_\cold^2 \rangle \rho_\cold + \langle g_\hot^2 \rangle \rho_\hot \right)^2}
T^2 X_2^2
+ \ldots
\, ,
\end{eqnarray}
and
\begin{eqnarray}
\langle \dot{Q}_\hot \rangle &=&
\frac{\kB T}{vI}
\rho_\cold \rho_\hot
	\frac{\langle g_\cold^3 \rangle \langle g_\hot^2 \rangle - \langle g_\cold^2 \rangle \langle g_\hot^3 \rangle}
	     {\left( \langle g_\cold^2 \rangle \rho_\cold + \langle g_\hot^2 \rangle \rho_\hot \right)^2}
T X_1
\nonumber \\&& \mbox{}
+ \frac{m}{M} v\kB T
\frac{\langle g_\cold^2 \rangle \langle g_\hot^2 \rangle \rho_\cold \rho_\hot}
	 {\langle g_\cold^2 \rangle \rho_\cold + \langle g_\hot^2 \rangle \rho_\hot}
T X_2
\nonumber \\&& \mbox{}
- \frac{\kB T}{4vI}
\rho_\cold \rho_\hot  T^2 X_1 X_2
\nonumber \\&& \mbox{} \times
\frac{R_I \left[\langle g_\cold^2 \rangle \rho_\cold \left( 15 \langle g_\cold^2 \rangle \langle g_\hot^3 \rangle
													- \langle g_\hot^2 \rangle \langle g_\cold^3 \rangle \right)
	- \langle g_\hot^2 \rangle \rho_\hot \left( 5 \langle g_\cold^2 \rangle \langle g_\hot^3 \rangle
													- 19 \langle g_\hot^2 \rangle \langle g_\cold^3 \rangle \right) \right]}
	 {\left( \langle g_\cold^2 \rangle \rho_\cold + \langle g_\hot^2 \rangle \rho_\hot \right)^3}
\nonumber \\&& \mbox{}
- \frac{1}{v m R_I^2}
	\frac{\langle g_\hot^2 \rangle \rho_\hot}
		 {\left( \langle g_\cold^2 \rangle \rho_\cold + \langle g_\hot^2 \rangle \rho_\hot \right)^2}
T^2 X_1^2
\nonumber \\&& \mbox{}
+ \frac{v \kB T}{4}\frac{m}{M}
\frac{\langle g_\cold^2 \rangle \langle g_\hot^2 \rangle \rho_\cold \rho_\hot
		\left( \langle g_\cold^2 \rangle \rho_\cold - \langle g_\hot^2 \rangle \rho_\hot \right)}
	 {\left( \langle g_\cold^2 \rangle \rho_\cold + \langle g_\hot^2 \rangle \rho_\hot \right)^2}
T^2 X_2^2
+ \ldots
\, ,
\end{eqnarray}
\endnumparts
where we defined the thermal velocity
\begin{equation}
    v = \sqrt{\frac{8 \kB T}{\pi m}}
\, .
\end{equation}
Conservation of energy
$\langle \dot{W}\rangle+ \langle \dot{Q_\hot}\rangle+\langle\dot{Q_\cold} \rangle =0$ follows immediately upon inspection.

The above expressions are the central result of this work. They express the first moments of thermodynamic quantities in the form of series expansion in both $\varepsilon$ and the thermodynamic forces $X_1$ and $X_2$, and incorporate in all detail the geometry of each part of the engine.  The approach taken here allows to go well beyond the thermodynamical linear regime. Although the algebraic calculations quickly become tedious and cumbersome, they are straightforward and easily achievable using a symbolic computation software tool.

Our approach extends the findings of \cite{chiralFridge}. In that work the authors considered an identical setup as shown in Figure~\ref{fig1} with the purpose to construct a chiral heat pump. In such a setup the external torque is used to drive a heat flow from the cold to the hot reservoir. The analysis of the device was done in the linear regime, making use of Onsager symmetry to complete the off-diagonal elements, without deriving them independently. However, as explained in \cite{chiralFridge},  the external torque not only induces a cooling flux, but also frictional heating in each reservoir. Since it is proportional to $\langle \omega \rangle \tau$, this frictional heating in fact is of second order in the torque, hence $\propto X_1^2$. A consistent in-depth analysis of the chiral pump inevitably must go beyond the linear regime. Our explicit results beyond the linear regine confirm the heuristic approach presented in \cite{chiralFridge}, but also reveal the appearance of additional non-linear terms in $X_1 X_2$ and $X_2^2$. 

As an application of our results, we focus on the efficiency of the microscopic Feynman ratchet. In the context of a heat engine, this efficiency is defined as the ratio of work over input heat,
\begin{equation}
    \eta = \frac{-\langle \dot{W} \rangle}{\langle \dot{Q_\hot} \rangle}.
\end{equation}
To see how this efficiency behaves upon varying the forces we write $\langle \dot{W} \rangle = a X_1^2 + b X_1 X_2$ and $\langle \dot{Q_\hot} \rangle = c X_1 + d X_2$, focusing for definiteness on the lowest orders. It is clear that $\langle \dot{W} \rangle$ equals zero for either $X_1=0$ (no external torque, and consequently no work) or $X_1^*=-(b/a)X_2$, also known as the stopping force. In this case the torque exactly cancels the steady angular velocity $\langle \omega \rangle =0$. In fact, only when $0 \leq X_1 \leq X_1^*$ does the setup behave as a heat engine. Keeping $X_2$ fixed, it is straightforward to prove that the delivered work is maximal when $X_1=X_1^*/2$ (see also \cite{PhysRevLett.95.190602,PhysRevLett.116.220601} and \cite{Tu_2008}). Setting $X_1=\lambda X_1^*$ we find for the efficiency:
\begin{equation}
    \eta =\frac{(1-\lambda )\lambda}{A - \lambda}T X_2
\end{equation}
with
\begin{equation}
    A=\frac{ad}{bc}=\frac{M}{m}\frac{8\gcc\ghh \left[\gcc \rho_\cold + \ghh \rho_\hot \right]^2 }{\pi\rho_\cold \rho_\hot \left[\gccc\ghh-\gcc \ghhh \right]^2}.
\end{equation}
Notice the presence of the factor $M/m=1/\ve^2$ implying $\alpha \gg \lambda$. Expressing $TX_2$ in terms of the Carnot efficiency $\eta_C$ yields
\begin{equation}
    TX_2=\frac{2\eta_C}{2-\eta_C}\approx \eta_C 
    \, , \quad
    \mbox{with }
    \eta_C=1-\frac{T_\cold}{T_\hot}
\, ,
\end{equation}
with the last approximation being valid for small temperature differences. We can conclude that the efficiency of the Feynman ratchet is just a small fraction of the Carnot efficiency (in the limit $\ve \ll 1$).

\section{Conclusion}\label{sectConc}
A unified framework is developed which treats the dynamical and thermodynamic variables of the microscopic Feynman ratchet on an equal footing. This allows for a detailed and systematic investigation of the thermodynamic performance. Explicit expressions are given for the lowest order moments of all the variables in the form of a series expansion in both $m/M$ and the thermodynamic forces. As a result it becomes possible to investigate the engine's performance in a systematic way beyond the linear regime.

\ack \addcontentsline{toc}{section}{Acknowledgments}
R.E. acknowledges financial support from the Swedish Research Council
(Vetenskapsr{\aa}det) under Grant No.~2020-05266. Nordita is partially supported by Nordforsk. B.C. acknowledges financial support from the Research Foundation - Flanders (FWO) under Grant No.~K226922N.

\section*{References}
\bibliography{feynman}

\end{document}